\documentclass[12pt]{article}

\usepackage{amssymb,graphicx}

\usepackage{epsf}
\usepackage{graphicx,epsfig}
\usepackage{amsfonts}
\usepackage{amssymb}
\usepackage{cite}

\def\br{{\bf r}}
\def\bx{{\bf x}}
\def\by{{\bf y}}

\def\mpl{M_{\rm Pl}}

\makeatletter
\renewcommand\section{\@startsection {section}{1}{\z@}%
                                 {-3.5ex \@plus -1ex \@minus -.2ex}%nn
                                   {2.3ex \@plus.2ex}%
                                   {\normalfont\large\bfseries}}
\renewcommand\subsection{\@startsection{subsection}{2}{\z@}%
                                   {-3.25ex\@plus -1ex \@minus -.2ex}%
                                     {1.5ex \@plus .2ex}%
                                     {\normalfont\bfseries}}
\renewcommand\subsubsection{\@startsection{subsubsection}{3}{\z@}%
                                   {-3.25ex\@plus -1ex \@minus -.2ex}%
                                     {1.5ex \@plus .2ex}%
                                     {\normalfont\itshape}}
\makeatother

\newcommand{\Letter}{
\setlength{\textwidth}{16.5cm}
   \setlength{\textheight}{22.6cm}
    \hoffset=-0.5in
\voffset=-2.1cm }

\Letter

\begin{document}
\newcommand{\be}{\begin{equation}}
\newcommand{\ee}{\end{equation}}
\newcommand{\bea}{\begin{eqnarray}}
\newcommand{\eea}{\end{eqnarray}}
\newcommand{\barr}{\begin{array}}
\newcommand{\earr}{\end{array}}

\thispagestyle{empty}
\begin{flushright}
\parbox[t]{1.5in}{MIT-CTP-4010}
\end{flushright}

\vspace*{0.3in}

\begin{center}
{\large \bf Decaying Hidden Dark Matter in Warped Compactification}

\vspace*{0.5in} {Xingang Chen}
\\[.3in]
{\em Center for Theoretical Physics 
\\ Massachusetts Institute of Technology, Cambridge, MA 02139 } 
\\[0.3in]
\end{center}

\begin{center}
{\bf Abstract}
\end{center}
\noindent
The recent PAMELA and ATIC/Fermi/HESS 
experiments have observed an excess of
electrons and positrons, but not anti-protons, in the high energy
cosmic rays. To explain this result, we construct a
decaying hidden dark matter model in
string theory compactification that incorporates the following two
ingredients, the hidden dark matter scenario in warped
compactification
and the phenomenological proposal of hidden light particles that
decay to the Standard Model. In this model, on higher dimensional
warped branes, various warped Kaluza-Klein
particles and the zero-mode of gauge field play roles of the
hidden dark matter or mediators to the Standard Model.

\vfill

\newpage
\setcounter{page}{1}

%\tableofcontents

%\newpage

\section{Introduction and model setup}
\setcounter{equation}{0}
\label{SecIntro}

More than eighty percent of the matter in the universe is dark
matter,
but its particle identities remains a mystery. Recent
experiments may have observed the first non-gravitational signals of
the dark matter. The PAMELA satellite
\cite{Adriani:2008zr} observed an excess
of cosmic ray positron fraction at energies from 10 GeV up to at least
100 GeV, but importantly not the anti-proton excess in the same energy
range. 
The ATIC balloon \cite{:2008zzr} found an excess of
cosmic
ray electrons above the astronomical background at energies 300-800 GeV
with a peak at around 600 GeV.
The recent Fermi satellite \cite{Abdo:2009zk}
also observed an excess of electrons in the
similar energy range, but with a much softer spectrum comparing to
ATIC. In addition, the HESS telescope \cite{Aharonian:2009ah}
observed a steepening of electron spectrum above TeV.
These electron or positron excess 
may be due to the annihilation or decay of the dark
matter.

Given the high energy scale of the events, the conventional dark
matter candidates such as the weakly interacting
massive particles (WIMP) would have annihilated or decayed into both
leptons and hadrons, including
electrons/positrons and protons/anti-protons
\cite{Cirelli:2008pk,Donato:2008jk}.
A possible
phenomenology behind this anomaly is 
that the final decay to the SM model particles is due
to a light particle
\cite{Finkbeiner:2007kk,Cholis:2008vb,ArkaniHamed:2008qn}. 
The mass of this light particle is
below about 1.8 GeV, so if decays on-shell the proton/anti-proton
production is kinematically forbidden. 
In order for this to work, the
annihilation channel or decay chain before that 
should be somewhat hidden from the
Standard Model (SM) sector. This suggests the hidden dark matter
scenarios
\cite{Chen:2006ni,Kikuchi:2007az,Harling:2008px,Gong:2008uz,Feng:2008ya,Feng:2008mu}.
In string theory compactification, sources carrying effective D3-brane
charges create warped throats in six extra-dimensions
\cite{Giddings:2001yu,Randall:1999ee,Verlinde:1999fy}. During
reheating, matter that is left in or tunnels to these throats can be
trapped by gravitational potentials. If the SM is located somewhere
else, these matter can become the hidden dark matter
\cite{Chen:2006ni}.
The main purpose of this
paper is to construct models that incorporate the above two ingredients.
We will combine the bottom-up and top-down approaches in constructing
such models.

There are large number of ways a hidden sector may be implemented in
terms of effective field theory. In string theory warped
compactification, a natural candidate is the hidden throat. 
We will use such configurations to motivate hidden dark matter models
that may or may not have simple 4d effective field theory
descriptions.
Various constraints, especially on the mediators between the
hidden and SM sector, can arise
from constructing models that have a reasonable UV completion.
These may not be easy to envision from 
the pure bottom-up approach.
On the other hand, experimental results play a crucial role in
specifying our model configuration. This gives the motivation to
explicitly construct some of the required components in string
configuration.

In this paper we would like to construct a decaying hidden dark matter
model in the warped compactification, which meets the following
model-building requirements:
\begin{enumerate}
\item The hidden dark matter is unstable and decays to the hidden
  light particles. This decay channel has a very long life
  time, such as of order $10^{26}$s. 

\item The hidden light particles decay on-shell to the SM particles
  in an astrophysically short lifetime, before they can escape the
  galactic halos, which has a size 10 kpc. This requires the lifetime
  to be smaller than $10^{12}$s.\footnote{I would like to thank Xuelei
  Chen and Xiaojun Bi for helpful discussions on this point.}

\item The direct decay of the hidden dark matter into the SM particles
  such as proton/anti-protons and electron/positrons is much slower
  than the above two channels combined.
\end{enumerate}

\begin{figure}
\begin{center}
%\epsfxsize=10cm
%\epsfbox{Fep.eps}
\epsfig{file=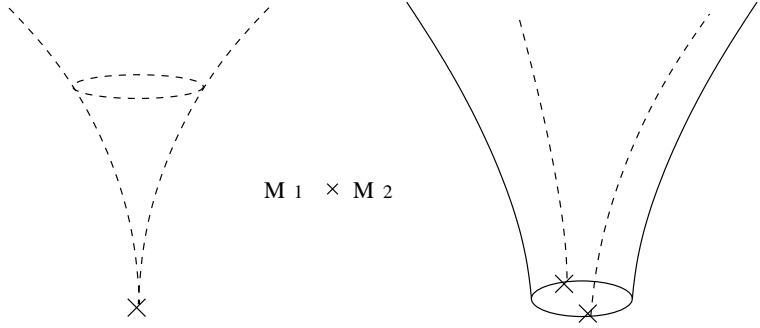, width=10cm}
\end{center}
\medskip
\caption{Configuration of the hidden throat. Crosses are the locations
  of the D3 or anti-D3-branes. Dashed lines
  indicate the higher dimensional branes; they wrap on the radial
  direction and the angular part $M_1$, and are embedded in $M_2$.}
\label{ThroatConfig}
\end{figure}

The following is our model configuration (Fig.~\ref{ThroatConfig}).
Consider a throat whose angular part is topologically a product,
$M_1 \times M_2$. The radius for $M_1$ shrinks to zero, while $M_2$
remains finite at the bottom of the throat. We add higher dimensional
spacetime-filling branes which wrap the radial direction and
$M_1$. They can also wrap part of $M_1$, as long as they share an
angular isometry in $M_1$.
The additional, if any, dimensions of these branes are embedded
in $M_2$. These higher dimensional branes extend outside the throat,
and intersect with or become part of the SM branes in the bulk.
Inside the throat, the gauge fields on the
higher dimensional branes
have warped Kaluza-Klein (WKK) excitations along the radial and $M_1$
directions; the $l$-th ($l>0$) partial wave WKK particles 
become our hidden dark matter. 
Graviton also has WKK excitations and can become part of
the hidden dark matter \cite{Chen:2006ni}.\footnote{If the SM is
  within the same throat,
  the stable gravity WKK particles \cite{Kofman:2005yz} can also
  become viable dark
  matter candidates \cite{Chen:2006ni}. They belong to the
  conventional WIMP.}
As we will see, comparing to the gravitons, 
the zero-modes or the $s$-wave WKK modes of
the gauge fields are much stronger mediators to the SM. The minimum
warped KK scale for this hidden throat is taken to be around
TeV, which is an input from the ATIC/Fermi/HESS. 
At the bottom of the throat,
there
are spacetime-filling D3-branes or anti-D3-branes.
Their locations can be separated in
$M_2$. The open strings on these D3-branes provide light fields
with masses hierarchically below TeV. This
gives the required hidden light particle. We take its mass to be
between 1.8 GeV and 1 MeV, so that it can decay to
electron/positron, but not proton/anti-proton. This is an
input from the PAMELA.

As mentioned, our model is motivated from both the existing
flux compactification in type IIB string theory \cite{Giddings:2001yu}
and the
experimental results. 
So on the one hand, the model will lead to a novel
field theory of the hidden dark matter;
on the other hand, not all components has been
constructed explicitly in the current string theory literature.
For example, before putting in
D3-banes and wrapping higher dimensional branes, the required throat
configuration is motivated from
the Klebanov-Strassler (KS) throat \cite{Klebanov:2000hb}, where the
topology
$M_1\times M_2$ that we described is $S^2\times S^3$. 
The higher dimensional branes can be D5 or D7-branes.
However, 
the specific configuration of the D3-branes, the wrapped and warped
D7-branes and their connection to the SM branes
are proposed to satisfy the ATIC and PAMELA experiments. The number of
the D7 and D3 branes 
should be small enough not to significantly affect the
original throat.

When estimating lifetimes of various fields, we will concentrate on
the most important factors regarding the above model building
requirements. 
For the purpose of this paper, it
suffices to ignore all the numerical pre-factors in our
approximations. 
Many other models on the decaying dark matter \cite{decayDMpaper}
or the annihilating dark matter \cite{anniDMpaper} 
have been studied in the context of the
PAMELA and ATIC. Possible astrophysical explanations are studied in
\cite{astroEx}.

\section{Gauge fields on warped brane}
\label{SecGauge}

We consider the example of wrapping D7-branes on the whole $S^2$ of
the KS throat.
The $l$-th ($l>0$) partial waves of the WKK modes have non-trivial
angular
dependence on $S^2$. When the size of $S^2$ shrinks to zero at the
tip of the throat, their wavefunctions have to vanish to avoid the
singularity, in contrast to the $s$-wave. 
So the wavefunctions do not overlap with the D3-branes located at
the
tip of the throat and these WKK modes are stable against decaying onto
the D3-branes, in contrast to the $s$-wave. In other words,
the presence of the
D3-branes at the tip of the throat do not break the $S^2$ isometry
associated with these partial waves, so we have conserved angular
momenta. 
Once we are aware of this
fact, we can ignore the very tip region of the throat and 
approximate the throat as
the following simple form to study the characteristic behavior of the
WKK wavefunction,
\bea
ds^2 = A^{-1/2} (-dt^2 + d\bx^2) + A^{1/2} (dr^2 + r^2 d\Omega_{S^2}^2
+ r^2 d\Omega_{S^3}^2) ~,
\label{warpmetric}
\eea
where
\bea
A=1+ \frac{R^4}{r^4} ~.
\label{warpfactor}
\eea
The $r$ is the radial coordinate of the AdS space and 
$R$ is the length scale of the warped space. The throat has a
cut-off at $r_0 \ll R$ and the minimum warp factor is $h_0 \equiv
r_0/R$.

The equation of motion for the Abelian gauge field 
$B^{(8)}_P$ with a zero 8d mass on D7-branes is
\bea
\partial_M \left( \sqrt{-G} G^{MP} G^{NQ} F^{(8)}_{PQ} \right) =0 ~,
\eea
where $F_{PQ}^{(8)} = \partial_P B^{(8)}_Q - \partial_Q B^{(8)}_P$ is
the field strength.
We decompose
$B^{(8)}_\mu 
\equiv {\tilde B}^{(4)}_\mu(x) \varphi(r) \Phi(\Omega_{S^2})$
and consider $B^{(8)}_m = {\rm const.}$. 
We use the Greek index $\mu$ to
label the $3+1$ spacetime dimensions, and the Roman index $m$ to label
the extra-dimensions. The tilde in ${\tilde B}^{(4)}_\mu$ indicates
that it has not been canonically normalized. 
$\Phi$ is the wavefunction on $S^2$, and the
wavefunction on another angular direction is taken to be
one. Choosing
the Lorentz gauge $\partial_\mu {\tilde B}^\mu_{(4)}=0$ 
and considering the
plane-wave ${\tilde B}^{(4)}_\mu \propto \epsilon_\mu e^{i p x}$, 
we get the
differential equation for $\varphi$ for the $l$-th partial wave in
$S^2$,
\bea
\frac{A^{1/4}}{r^2} \frac{d}{dr} \left( A^{-1/4} r^2 \frac{d}{dr}
\varphi \right) - \frac{l(l+1)}{r^2} \varphi + m^2 A \varphi =0 ~,
\label{eom_varphi}
\eea
where $m^2 = -p^2$ is the four-dimensional mass of the particle.

We are interested in the particles that are trapped
inside the throat, so we impose the outgoing boundary condition. 
Solving (\ref{eom_varphi}), we get the following spectrum. It consists
of a massless zero-mode, whose wavefunction on the D7-branes are
constant, and a tower of WKK particles with
masses quantized in unit of $\sim h_0 R^{-1}$. 
At each level of the WKK, we have
different partial waves labeled by $l$, and their wavefunction in the
radial direction behave as
\bea
\varphi \propto r^{-b_l -1} ~,
~~~~~~ b_l = \sqrt{l^2+l+1} ~,
\label{varphi_beh}
\eea
for $m R^2 \ll r \ll R$; $\varphi \propto r^{-l-1}$ for $R\ll r \ll
m^{-1}$. Comparing to the gravity modes, for the zero-mode, the
world-volume of
D7-branes on which the gauge
fields propagate is only a much smaller subspace of the whole bulk; for
the WKK modes, the damping of the wavefunction is
much slower than that of the gravity which is for example
$r^{-4}$ for the $s$-wave \cite{Chen:2006ni}. These facts will become
important later.
The zero-mode disappears if the gauge field acquires a 8d mass.

\section{Dark matter decay within hidden throat}
\label{SecDMdecay}

In a realistic compactification, the angular isometries of the throat
will always be broken, for example at least 
by the presence of the SM branes
outside the throat. Let us consider the effect of a stack of
$\tilde N$ extra
D3-branes located at $\br=\by_0$. 
Instead of (\ref{warpfactor}), the warped geometry is now a
superposition of that of the throat and branes,
\bea
A &=& 1+ \frac{R^4}{r^4} + \frac{{\tilde N} R^4}{ N |\br-\by_0|^4} ~,
\nonumber \\
&=& 1+ \frac{R^4}{r^4} + \frac{ {\tilde N} R^4}{N y_0^4}
\left( 1+ \frac{4r}{y_0} \cos \theta + \cdots \right) ~,
\label{deformedA}
\eea
where we have expanded around $r=0$ for $r\ll y_0$, $\theta$ is
the angle between $\br$ and $\by_0$, $N$ is the charge of the
hidden throat.

The angular dependence in (\ref{deformedA}) introduces a mixing
between the $l$-th ($l \ne 0$) and $s$-wave WKK mode in the following
otherwise vanishing kinetic term,
\bea
\sim 
\frac{m_s^4}{g_s} \int d^8X \sqrt{-G_{(8)}} F^{(8)}_{MN} F^{(8)}_{PQ}
G_{(8)}^{MP} G_{(8)}^{NQ} ~,
\label{Bkinetic}
\eea
where one of $F^{(8)}_{MN}$ is for the $l$-th
partial wave and another the $s$-wave.
Integrating the extra dimensions using the wavefunction
(\ref{varphi_beh}) and the deformed metric (\ref{deformedA}), and
normalizing the four-dimensional fields ${\tilde B}^{(4)}_\mu$ using
(\ref{Bkinetic}) for the $s$-wave and $l$-th partial wave
respectively, we get the mixing
\bea
\epsilon \sim h_0^5 \frac{\tilde N}{N} \left( \frac{R}{y_0} \right)^5
~.
\label{ls-mixing}
\eea
Due to the wavefunction (\ref{varphi_beh}), the integration is
dominated by the contribution from the tip of the throat for $l\ge
2$. The parameter (\ref{ls-mixing}) is predominant over the
other model-dependent parameters that we will assume on the hidden
particle model on D3-branes.

On the D3-branes at the tip of the throat, there are light
particles. The 8d $s$-wave WKK gauge field intersects with the
D3-branes and 
induces a 4d gauge field
through a mixing strength $\alpha_1$, which is determined by the
string coupling $g_s$ between the open strings on D7-branes and
D3-branes. The 4d gauge field couples to other light
particles, for example, the fermion-anti-fermion $\psi/\bar \psi$ with
a coupling constant $e_1$. Neither $\alpha_1$ nor $e_1$ depends on the
warp factor. So the coupling of the $s$-wave
$B_\mu^{(8)}$ to $\psi/\bar \psi$ is given by
\bea
e_1 \alpha_1 \int d^4 x \bar \psi \gamma^\mu B_\mu^{(8)} \psi ~,
\label{Bpsipsi}
\eea
where $B_\mu^{(8)}$ is evaluated at the location of the D3-branes.

Working with the canonically normalized fields $B^{(4)}_{\mu}$, 
using (\ref{ls-mixing}) and (\ref{Bpsipsi}), we get the decay
rate of the $l$-th WKK to two $\psi$'s,
\bea
\Gamma_{l\to s \to 2\psi} \sim h_0^{10} 
\left( \frac{R}{y_0} \right)^{10} 
\frac{\tilde N^2}{N^3} ~e_1^2 \alpha_1^2~
 m_{\rm WKK} ~.
\label{lto2psi_s}
\eea
We have used the relation $m_{\rm WKK} \sim h_0 R^{-1}$ and $R
\sim (g_s N)^{1/4} m_s^{-1}$. The fermions can quickly cascade into
many lighter particles on D3-branes, such as a neutral boson $\chi$
that is stable before decaying outside the throat.

When we integrate the extra-dimensions in (\ref{Bkinetic}), the
integrand is proportional to $r^{-b_l+1}$. For $l\ge 2$, the main
contribution comes from the tip region. This is what we used
above. For $l=1$, the integration gets most contribution toward the
UV region, and we integrate it up to the location of the extra branes,
$\by_0$. The mixing becomes larger. In fact, for low $l$, there is
another channel that can give a competitive decay rate. That is,
the $l$-th WKK can also decay through mixing with the zero-mode gauge
field. Because the zero-mode has a constant wavefunction, in the
integration over the extra dimensions, the integrand is less peaked at
the tip of the throat. This integration gets most contribution near
the UV region for $l \le 3$. 
The zero-mode gauge field couples to $\psi/\bar \psi$ through the same
term (\ref{Bpsipsi}).
The decay rate through this channel is
\bea
\Gamma_{l\to 0 \to 2\psi} \sim
h_0^{2b_l+2} \left( \frac{R}{y_0} \right)^{2b_l+2}
\left( \frac{R}{L_b} \right)^8 \frac{\tilde N^2}{N^3}
~e_1^2 \alpha_1^2~ m_{\rm WKK} ~, 
~~~~~ l \le 3 ~.
\label{lto2psi_0}
\eea
Although the power on the warp factor is smaller than (\ref{lto2psi_s}),
(\ref{lto2psi_0}) gets an extra suppression from the D7-brane
extra-dimensional volume $L_b^4$ 
because $L_b$ is larger than the size of the throat
$R$. As we
will discuss in Sec.~\ref{SecDMdirect}, this is a channel we
want to avoid.
For $l \ge 4$, a calculation similar to (\ref{lto2psi_s}) shows that the
decay rate through this channel
is smaller than (\ref{lto2psi_s}) by a factor of
$(R/L_b)^8$, so it is always highly suppressed.

To give a numerical example, suppose
$e_1 \sim 0.1$, $\alpha_1 \sim 0.1$, $\tilde N \sim 10$,
$N\sim 100$ and $y_0
\sim R$. The mass of the WKK is determined by the ATIC/Fermi/HESS 
to be around
TeV. In order for the lifetime of the dark matter to be
$10^{26}$s,\footnote{The lifetime of the dark matter may be shorter
  if the light particles which eventually decay to electrons/positrons
  have a small production branching ratio here.}
from (\ref{lto2psi_s}) we get the warp factor of the hidden
throat, $h_0 \sim 10^{-4.5}$. It is exponentially small, as naturally
realized in the flux compactification.

It is clear
that in this scenario the decay rate is highly dependent of the
environment, such as the warp
factor of the hidden throat and the location of the extra branes.
This is also seen for the gravity WKK
modes as the conventional WIMP
\cite{Berndsen:2008my,Dufaux:2008br,ClineProgress}.
For the gravity WKK modes, it is possible
to systematically classify different perturbations that break the
angular isometries
\cite{Berndsen:2008my,Dufaux:2008br,ClineProgress,Ceresole:1999ht}. 
It may be done similarly here, although the coefficients
of the classifications still need to be determined by a detailed
configuration. We have
used a simple configuration to convey the following
main message regarding the requirement 1 in Sec.~\ref{SecIntro}.
The isometry breaking, that is present in the bulk or
UV side of the throat, is separated from the IR tip by the
gravitational potential of the warped geometry, so
the decay rate is typically suppressed
by a large power of the warp factor. Hence the lifetime of such a
TeV scale hidden
dark matter is very long. However the precise decay rate
and the energy scale of the cosmic rays are typically not the specific
predictions of general hidden dark matter models \cite{Chen:2006ni}. 
In fact, it is natural in this
scenario that there are multiple peaks of cosmic rays
at various energy scales and associated with
different dark matter lifetimes. 
We are just starting to observe one of them.
We now turn to the next two requirements concerning the various
communications between the hidden throat and the SM.

\section{Decay of hidden light particle to SM}
\label{SecDecayLP}

\begin{figure}
\begin{center}
%\epsfxsize=10cm
%\epsfbox{Fep.eps}
\epsfig{file=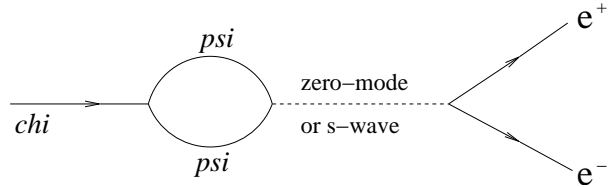, width=8cm}
\end{center}
\medskip
\caption{Feynman diagram for the decay of the light particle $\chi$ to
electron/positron.}
\label{Fdiagram}
\end{figure}

The decay of the hidden light particle $\chi$ to the SM
electron/positron can
be mediated by the gauge fields on the D7-branes
because they extend outside the throat and intersect with the SM
located at a distance $D$. 
The zero-mode and $s$-wave are most important for this purpose, since
higher partial waves damp faster. A Feynman diagram of such a decay is
given in Fig.~\ref{Fdiagram}.

We first consider the zero-mode.
The particle in the loop is a hidden fermion that 
couples to the hidden D3-brane gauge
field. As mentioned in Sec.~\ref{SecDMdecay}, this induces a coupling
to the D7-brane gauge field, and in turn induces a coupling to the SM
electro-magnetic gauge field.
The first vertex is given by a Yukawa coupling $\lambda_1$.
Both the second and third vertex are given by the term
(\ref{Bpsipsi}), at different locations and 
with different couplings $e_{1,2}$,
$\alpha_{1,2}$ and particle contents $\psi$, $\psi_e$. 
In terms of the normalized 4d field $B^{(4)}_\mu$,
both couplings receive a suppression factor from the D7-brane volume,
$(R/L_b)^2 N^{-1/2}$.
This is the most important parameter for this process.
Comparing to the $s$-wave mediation that we will study shortly, 
there is no warp factor here 
because the wavefunction of the zero-mode is
constant on D7-branes.
Comparing to the graviton zero-mode, which effectively gives a
coupling $m_{\rm WKK}/\mpl \sim h_0 (R/L)^3 N^{-1}$ 
($L$ is the size
of the bulk), the coupling here is much stronger for the following
reasons. First, the D7-branes are
embedded in the bulk, so its size is typically smaller, $L_b<L$;
second, the D7-brane has fewer dimensions to integrate over; 
third, the
couplings $e_{1,2}$, $\alpha_{1,2}$ are already dimensionless so it is
not affected by warping.

The decay rate of $\chi$ through this channel is
\bea
\Gamma_{\chi\to0\to2e} \sim
\left( \frac{R}{L_b} \right)^8 \frac{1}{N^2}
~\lambda_1^2 e_1^2 e_2^2 \alpha_1^2 \alpha_2^2 ~
\frac{\Lambda^4}{m_\chi^4} \frac{E_0}{\gamma} ~.
\label{chito2e_0}
\eea
The $\Lambda$ is the momentum cutoff in the loop of
$\psi$, as an example 
we approximate it to be $m_{\rm WKK}\sim {\rm TeV}$ which is
the warped KK scale at the bottom of the throat. 
We assume $m_\chi \gg 2m_e$. The energy scale in the center-of-mass
frame is $E_0 \sim m_\chi$, and $\gamma \sim m_{\rm WKK}/m_\chi$ is the
Lorentz factor for $\chi$. The denominator
$m_\chi^4$ comes because the
zero-mode mediator is massless. So if there are
several stable light particles as $\chi$, the lightest one is most
favorable to decay in this channel.

For the $s$-wave mediation, the most important factor is the suppression
factor in the third vertex, $h_0^2 (R/D)$ for $D>R$ and $h_0^2(R/D)^2$
for $D<R$, due to the WKK damping outside the throat.
The decay rate through this channel is
\bea
\Gamma_{\chi\to s \to 2e} \sim
h_0^4 \left( \frac{R}{D} \right)^2 \frac{1}{N^2}
~\lambda_1^2 e_1^2 e_2^2 \alpha_1^2 \alpha_2^2 ~
\frac{\Lambda^4}{m_{\rm WKK}^4} \frac{E_0}{\gamma} ~.
\label{chito2e_s}
\eea
The factor $(R/D)^2$ becomes $(R/D)^4$ for $D<R$. The heavier $\chi$ is
more favorable to decay in this channel.

Let us use the same numerical example in Sec.~\ref{SecDMdecay} and
assume $\lambda_1\sim e_2 \sim \alpha_2 \sim 0.1$. The $R/L_b$ can be
as small as $R/L \sim (N h_0^{-1} m_{\rm WKK}/\mpl)^{1/3} \sim
10^{-3}$, so we use $R/L_b \sim 10^{-2}$ for example. The decay rate
$\Gamma_{\chi\to 0 \to 2e}$ then ranges from $10^3$ to $10^9{\rm
  s}^{-1}$, for $m_\chi$ from GeV to MeV. The decay mediated by the
$s$-wave is much slower, $\Gamma_{\chi \to s \to 2e} \sim 10^{-11}$ to
$10^{-17}{\rm
  s}^{-1}$. For some parameter space this lifetime is still short
enough to satisfy our requirement.
This channel becomes important in case the zero-mode gets lifted.
If we instead use the $s$-wave gravity WKK
\cite{Chen:2006ni} to mediate the decay, we will have an extra factor
of $h_0^4$. This makes the lifetime of $\chi$ easily exceed the age
of the universe.\footnote{The case where both the hidden
  dark matter and hidden light particle have lifetimes longer than
  the age of the
  universe is a viable case, but the decay products are no longer
  observable since they escape the galactic halos.}

\section{Direct decay of dark matter to SM}
\label{SecDMdirect}

A direct decay of the hidden dark matter to the SM should produce both
electrons/positrons and
protons/anti-protons, among many other particles. A rate larger
than (\ref{lto2psi_s}) would generically contradict the PAMELA
experiment.
There are a couple of ways a direct decay can happen.

First, since the wavefunction of the WKK dark matter itself has a
damping tail 
outside the throat,
the direct decay can be mediated by the term
(\ref{Bpsipsi}), but now we use the $l$-th partial wave in
$B^{(8)}_\mu$ and evaluate it at the location of the SM branes. 
The decay rate is
\bea
\Gamma_{l \to 2e} \sim  
h_0^{2b_l +2} \left( \frac{R}{D} \right)^{2l+2} 
\frac{1}{N} ~e_2^2 \alpha_2^2 ~ m_{\rm WKK} ~.
\label{lto2e}
\eea
The factor $(R/D)^{2l+2}$ becomes $(R/D)^{2b_l +2}$ for $D<R$.
The decay rate to the proton/anti-proton or other hadrons is the same
except that the couplings $e_2$ and $\alpha_2$ may be different. 
Eq.~(\ref{lto2e}) decreases drastically as $l$ increases, because
larger angular momentum introduces
higher effective potential. Comparing with (\ref{lto2psi_s}),
for $l \ge 4$, (\ref{lto2e}) has
at least additional powers of warp factor and is highly
suppressed. So all requirements in Sec.~\ref{SecIntro} are naturally
satisfied.  For $l\le 3$, (\ref{lto2e}) has a smaller power on
$h_0$. In order to suppress (\ref{lto2e}), 
one can decrease $y_0$ to increase the
isometry-breaking and increase $D$ to move away the SM branes. Neither
case affects the $\chi$ decay rate (\ref{chito2e_0}).
(But both affect (\ref{chito2e_s}).)
In this case, we are introducing isometry breaking sources besides
the SM branes. For the most difficult case, $l=1$, using
(\ref{chito2e_0}) it is enough to
have, for
example, $y_0 \sim 10^{-3} R$, $D\sim 10^2 R$ and $h_0 \sim
10^{-7.5}$.

Second, the direct decay can be mediated by mixing with 
virtual particles. 
The channel through a
virtual $s$-wave WKK is doubly suppressed by both mixing and
tunneling. For the virtual
zero-mode, we need to suppress the decay rate (\ref{lto2psi_0})
relative to (\ref{lto2psi_s}). This is because the zero-mode gauge
field couples similarly to the D3-branes both 
inside and outside the throat
through the term (\ref{Bpsipsi}), so this channel mediates the hadron
production. Again, this suppression is naturally achieved for $l\ge
4$. For $l \le 3$, especially $l=1$,
we need to significantly increase the D7-brane volume
suppression. In our previous example, $R/L_b$ can be as small as
$10^{-3}$, which is enough for the suppression to work for all $l\le
3$.

Alternatively, the absence of the first few low-$l$ ($l>0$)
partial waves can be achieved
by some discrete symmetries in the angular directions. For
example, a discrete symmetry on the $S^2$ azimuthal angle under 
$\varphi \to \pi/2 + \varphi$ makes the partial wave start from
$l=4$.

\section{Summary and discussion}
\label{SecSum}

In this paper we have constructed a decaying hidden dark matter model
in warped compactification. Generally speaking, in this class of
models,
there is an angular isometry shared by the hidden throat and wrapped
higher dimensional branes, and this isometry is not broken by the
presence of the
D3-branes (or anti-D3-branes) at the tip of the throat.\footnote{This
  requirement is necessary only if the hidden dark matter is the WKK
  modes discussed in Sec.~\ref{SecGauge} \& \ref{SecDMdecay}. It can
  be dropped if the hidden dark matter is something else in the hidden
particle sector (e.g.~analogous to the lightest supersymmetric
particles for SM). 
In the latter case, the only requirement for the higher
dimensional branes is that they intersect (or nearly intersect) 
both the SM and
hidden particle sectors. The mediation between the two sectors 
is provided by the
zero-mode gauge field on the higher dimensional branes.}
There is a mass hierarchy between
the WKK modes (of effective fields from open or closed
strings) associated with this
isometry and the light particles on the D3-branes, which serve as the
hidden dark matter and the hidden light particles, respectively. 
The communication
between the hidden throat and the SM sector is mediated by the
zero-mode or the $s$-wave WKK mode of the gauge field (or other
similar fields) on the higher dimensional branes.\footnote{By
  construction, the mediator here is electro-magnetically neutral and
  conserves the SM lepton and hadron number. So the stable SM
  particles do not decay through this mediator.} 

\begin{figure}
\begin{center}
%\epsfxsize=10cm
%\epsfbox{Fep.eps}
\epsfig{file=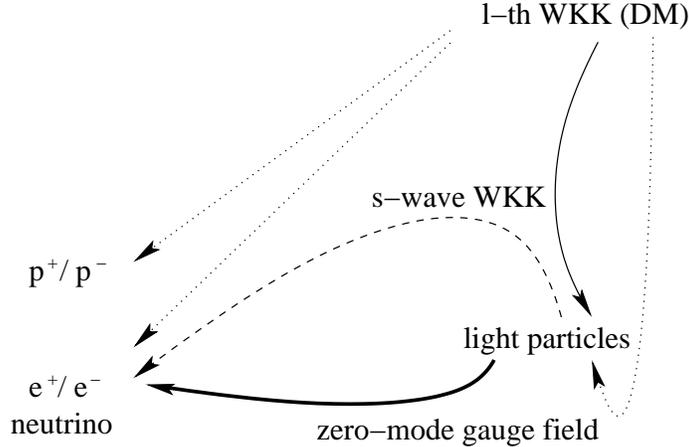, width=9cm}
\end{center}
\medskip
\caption{The decay channels of the WKK dark matter. The solid lines
  are the dominant decay channels. The dashed line is a much slower
  channel. The dotted lines are highly suppressed. Curved lines
  indicate the virtual particles involved in the Feynman diagrams.}
\label{DecayChannel}
\end{figure}

We studied an example of such a model and found the following main
features (Fig.~\ref{DecayChannel}).

\begin{enumerate}
\item The objects that break the isometry associated with the WKK
  modes are
  separated by the gravitational potential of the throat from the WKK
  particles. So the hidden dark matter, either gauge field or gravity
  WKK particles, have very long lifetime.

\item The decay of light particles to SM electrons/positrons can be
  mediated by the zero-mode or the lowest $s$-wave WKK mode of the
  gauge field on the higher dimensional branes. 
  The zero-mode is efficient (especially
  for interaction terms with dimensionless coupling
  constants), because its wavefunction does not damp as WKK, and the
  volume suppression is much weaker than the gravity zero-mode. This
  mediation mechanism is very important for the communication between
  the hidden sector and SM.
  In case the zero-mode gets lifted, the $s$-wave 
  mediation may also be sufficient (although much weaker), because its
  wavefunction damps much slower than the gravity WKK.

\item The direct decay of the dark matter to the SM particles is
  highly suppressed. This 
  is because the direct tunneling is suppressed by the effective
  potentials caused by the warped space and WKK's angular momenta;
  while
  the channel through the virtual gauge field zero-mode 
  (or $s$-wave WKK)
  is suppressed
  by both the small mixing and the volume of the higher dimensional
  branes (or the tunneling).
\end{enumerate}

Many aspects are worth to be further studied.
In a realistic warped
compactification, there are other components and fields
in the hidden throat
besides what we have considered.
It will be interesting to explore their roles in
this model.
It is also interesting to construct the required brane
configuration more explicitly in string theory.
At least, any SM particles lighter than the electron, i.e.~neutrino
and photon, are kinematically allowed in the decay, but can have
different branching ratios;
on the hidden 
D3-branes, there can also naturally be massless particles.
The associated observational effects such as the high energy gamma ray
or neutrino emissions is an important issue.
It is also important to study possible
collider physics signals.

\medskip
\section*{Acknowledgments}

I would like to thank Xiaojun Bi, Xuelei Chen, Alejandro Jenkins, Igor
Klebanov and
Henry Tye for helpful
discussions and comments. This
work is supported by the US Department of Energy under cooperative
research agreement DEFG02-05ER41360.

%\appendix

%\section{Appendix}

\end{document}